\documentclass[a4paper]{spie} 

\usepackage[]{graphicx}
\usepackage{aas_macros}
\usepackage[]{subfigure}

\title{A very compact  polarizer for an X-ray  polarimeter calibration} 

\author{Fabio~Muleri\supit{a,b}, Paolo~Soffitta\supit{a}, Ronaldo~Bellazzini\supit{c}, Alessandro~Brez\supit{c}, Enrico~Costa\supit{a}, Sergio~Fabiani\supit{a}, Massimo~Frutti\supit{a}, Massimo~Minuti\supit{c}, Maria~Barbara~Negri\supit{d}, Piermarco~Pascale\supit{a}, Alda~Rubini\supit{a}, Giuseppe~Sindoni\supit{a}, Gloria~Spandre\supit{c}
\skiplinehalf
\supit{a} Istituto di Astrofisica Spaziale e Fisica Cosmica, Via del Fosso del Cavaliere 100, I-00133 Roma, Italy;
\\
\supit{b} Universit\`{a} di Roma Tor Vergata, Dipartimento di Fisica, via della Ricerca Scientifica 1, 00133 Roma, Italy
\\
\supit{c} Istituto Nazionale di Fisica Nucleare, Largo B. Pontecorvo 3, I-56127 Pisa,  Italy
\\
\supit{d} ASI, Agenzia Spaziale Italiana, Viale Liegi 26, I-00198 Roma, Italy
}

\authorinfo{Further author information: (Send correspondence to Fabio Muleri)
\\Fabio Muleri: E-mail: Fabio.Muleri@iasf-roma.inaf.it,
Telephone: +39-0649934565}

\begin{document} 
\maketitle 

\begin{abstract}
We devised and built a light, compact and transportable X-ray polarized source based on the Bragg diffraction at nearly 45 degrees.  The source is composed by a crystal coupled to a small power X-ray tube. The angles of incidence  are selected by means of two orthogonal capillary plates which, due to the small diameter holes (10 $\mu$m) allow good collimation with limited sizes. All the orders of diffraction defined by the crystal lattice spacing are polarized up to the maximum order limited by the X-ray  tube voltage. Selecting suitably the crystal and the X-ray tube, either the line or the continuum emission can be diffracted, producing polarized photons at different energies. A very high degree of polarization and reasonable fluxes can be reached with a suitable choice of  the capillary plates collimation. 

We present the source and test its performances with the production of nearly completely polarized radiation at 2.6, 5.2, 3.7 and 7.4~keV thanks to the employment of graphite and aluminum crystals, with copper and calcium X-ray tubes respectively. Triggered by the very compact design of the source, we also present a feasibility study for an on-board polarized source, coupled to a radioactive Fe$^{55}$ nuclide and a PVC thin film, for the calibration of the next generation space-borne X-ray polarimeters at 2.6 and 5.9 keV. 
\end{abstract}


\section{Introduction}

The on-going development of new generation X-ray polarimeters based on photoelectric absorption will allow the study of polarized emission from astrophysical sources in some years, after about 35 years of wait. In particular the very compact Gas Pixel Detector\cite{Costa2001,Bellazzini2006,Bellazzini2006c} will perform polarimetry of galactic and extragalactic sources in a few days \cite{Costa2006,Bellazzini2006b}, allowing to study in deeper details a great variety of X-ray sources.

While the Gas Pixel Detector allows a large increase of polarimetric sensitivity respect to the X-ray polarimeters flown to date, it is based on a new-generation pixel detector and hence it requires for the first time the study of the response to polarized and unpolarized radiation in the whole energy range of interest. Since we are currently planning its use with a DME and Helium based mixture between $\sim$2 and $\sim$10~keV, to exploit both the X-ray optics band-pass and the higher flux from astrophysical sources at low energy, we need a polarized source in this energy range. Moreover the polarimetric sensitivity, estimated with a Monte Carlo software, peaks at energy of about 3~keV and it is rapidly changing with energy, strongly demanding a polarized source at energy of some keV.

Synchrotron facilities can readily produce polarized radiation in this energy range, but they can hardly be employed for the continuous and time-consuming development of a new detector, which requires a compact and readily available calibration source. Hence, up to now, we were forced to drive the detector development with a Monte Carlo software and with sensitivity measurements at 5.4 and, occasionally, at 8.04~keV. Polarized photons were produced by means of 90$^\circ$ Thomson scattering on a lithium target, enclosed in a thin beryllium case to prevent its oxidation and nitridation. The employment of this source is limited at low energy by the photoelectric absorption in the lithium target and in the beryllium case, which results in extremely merge fluxes at energy of a few keV.

In this paper we present a new polarized source, designed for the calibration of the Gas Pixel Detector in the energy range of maximum sensitivity, based on the Bragg diffraction at 45$^\circ$. Selecting properly the X-ray tube employed as unpolarized photons source, the diffracting crystal and the collimators that constrain the diffraction angle, polarized photons at 2.6, 5.2, 3.7 and 7.4~keV can be produced with reasonable fluxes and very high degree of polarization.

The very compact design of this source suggests its use as on-board calibration source, with unpolarized incoming photons produced with a radioactive source. We present a design and a feasibility test which suggests that a source of polarized photons at 2.6 and 5.9 keV, covering the energy range where the maximum sensitivity of the Gas Pixel Detector is reached, can be built with just a moderately strong Fe$^{55}$ radioactive source and exploiting fluorescence from chlorine atoms.

In sec.~\ref{sec:Bragg} and \ref{sec:Source} we respectively describe how Bragg diffraction can be used to build a polarized source and the source construction, while the spectrum of output radiation and the estimated polarization are reported in sec.~\ref{sec:Performances}. In sec.~\ref{sec:OnboardSource} we present a concept design of the on-board calibration source with the feasibility test we made.

\section{The Bragg diffraction at 45$^\circ$} \label{sec:Bragg}

Bragg diffraction, involving the diffraction of photons on the planes of a crystal lattice, is sensitive to the polarization of incident radiation. If $P_\lambda(\theta)$ is the diffracted intensity when parallel monochromatic radiation of unit intensity is incident at glancing angle $\theta$ to the diffracting planes of a flat crystal, then $P_\lambda(\theta)$ is strongly peaked at the angle which satisfies the Bragg condition. The energy,  within few eV for a flat crystal, and the angle of diffraction are hence related by the relation:
\begin{equation}
E = \frac{nhc}{2d\sin\theta}, \label{eq:BraggLaw}
\end{equation}
where $h$ and $c$ are respectively Planck's constant and the speed of light, $d$ the crystal lattice spacing and $n$ the diffraction order.

The efficiency of diffraction is expressed with the integrated reflectivity $R_\lambda$, defined as \cite{Evans1977}:
\begin{equation}
R_\lambda = \int_0^\frac{\pi}{2} P_\lambda \left(\theta\right) d\theta.
\end{equation}

\begin{figure}[htbp]
\begin{center}
\begin{tabular}{c}
\subfigure[\label{fig:k}]{\includegraphics[angle=0,totalheight=4.5cm]{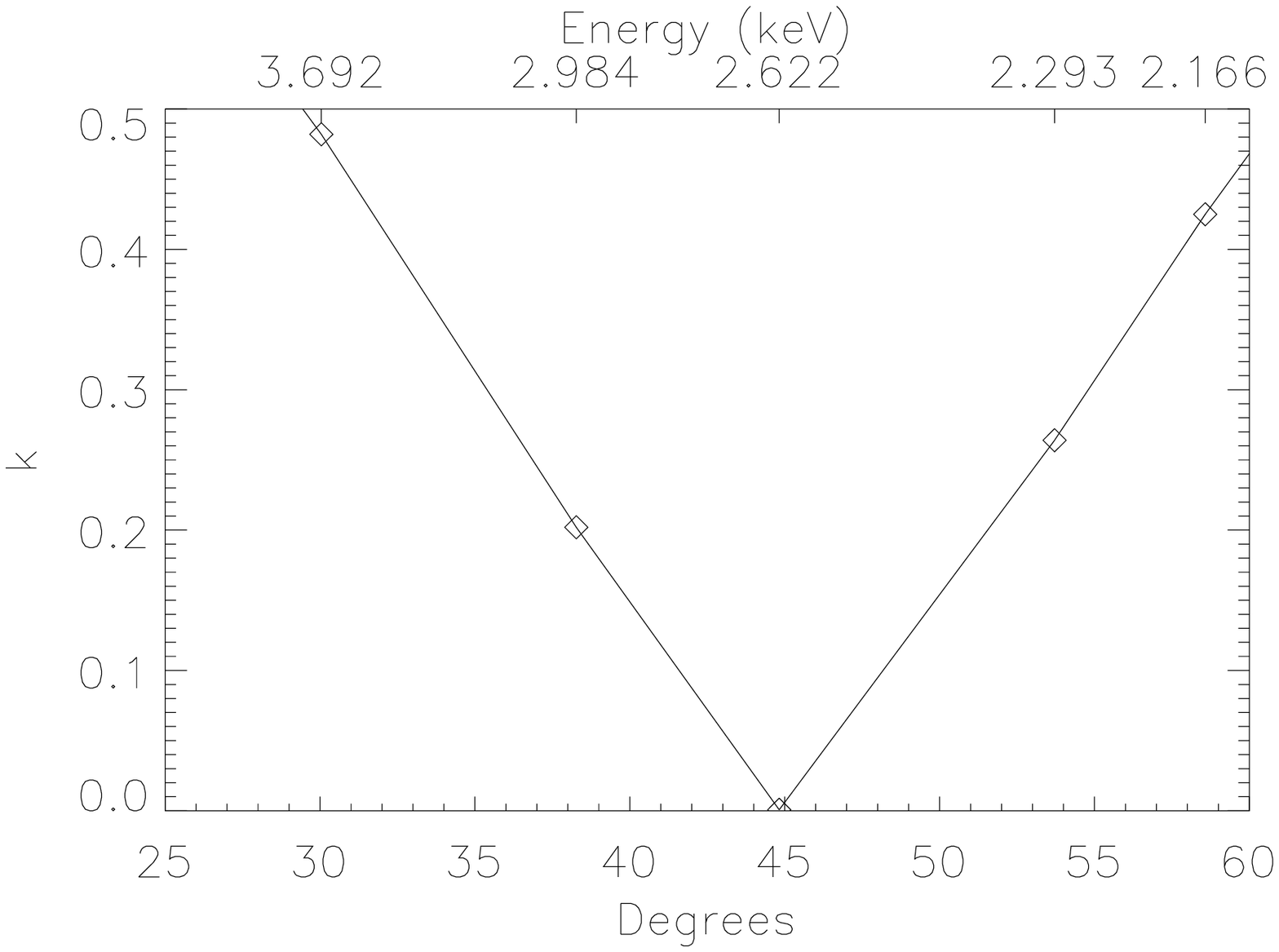}}
\subfigure[\label{fig:P}]{\includegraphics[angle=0,totalheight=4.5cm]{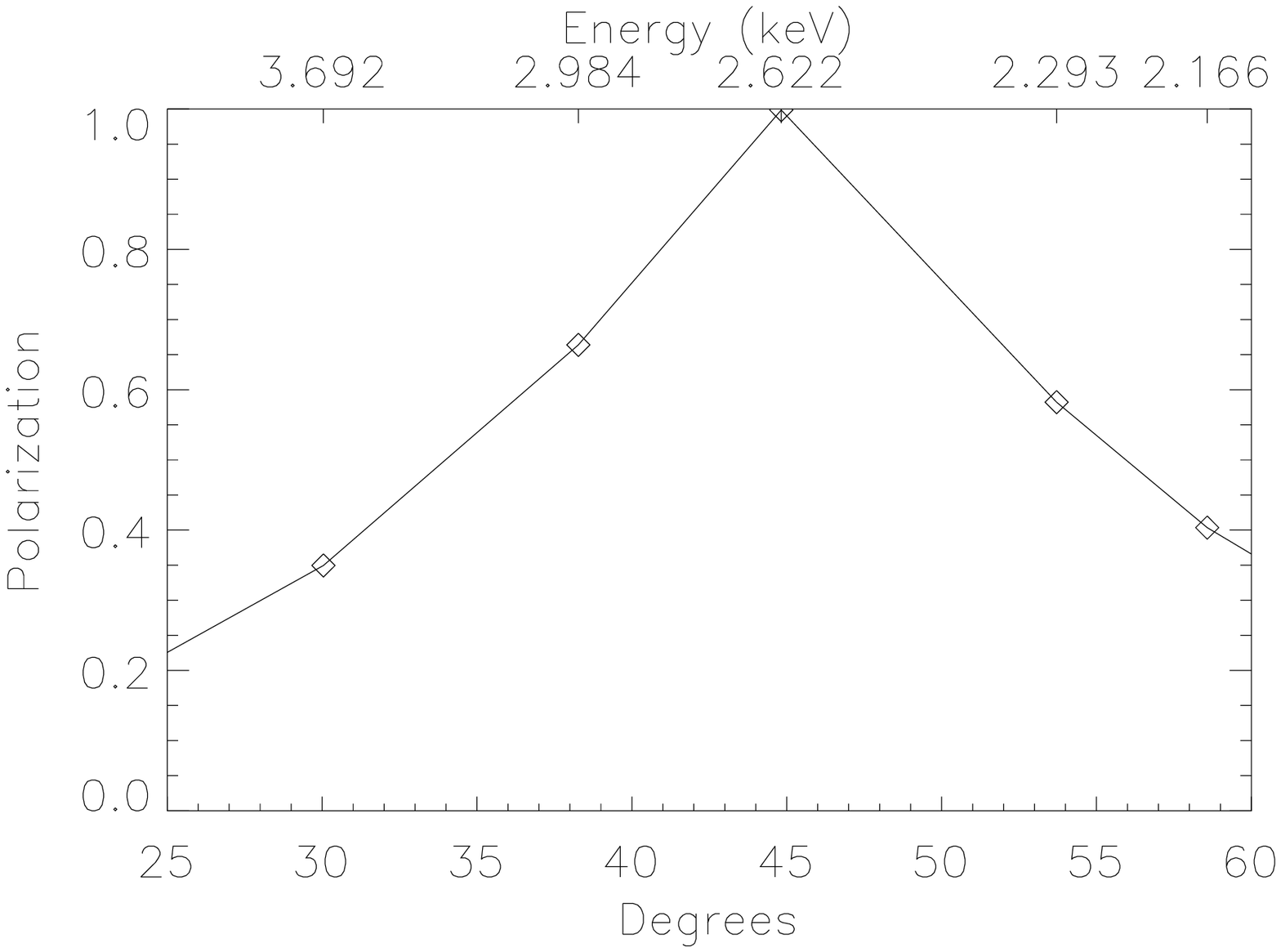}}
\end{tabular}
\end{center}
\caption{\small ({\bf a}) Ratio  $k={R_\lambda^\pi}/{R_\lambda^\sigma}$ of integrated reflectivity for incident radiation completely polarized parallel and perpendicular to the incidence plane for a graphite crystal. $k = 0$ for diffraction at 45$^\circ$ (data from Henke et al.\cite{Henke1993}). ({\bf b}) Degree of polarization for diffracted radiation from a graphite crystal, calculated with eq.~\ref{eq:PolarizationDegree} (data from Henke et al.\cite{Henke1993}). Diffraction angle and energy are related by Bragg's Law.}
\end{figure}

The integrated reflectivity depends on the polarization direction of incident radiation. Defining the plane of incidence as the plane where the direction of incident radiation and the normal of diffracting planes lie, the polarization of radiation can be decomposed in a component parallel ($\pi$-component) and perpendicular ($\sigma$-component) to the plane of incidence. The ratio $k$ of integrated reflectivity of the two components is plotted in fig.~\ref{fig:k} for a graphite crystal. When $k<1$, the diffracted radiation is partially polarized in the direction perpendicular to the incidence plane and the degree of polarization is \cite{Evans1977}:
\begin{equation}
{\cal P} = \frac{1-k}{1+k}. \label{eq:PolarizationDegree}
\end{equation}

In fig.~\ref{fig:P} we report the degree of polarization of radiation diffracted from a graphite crystal. When the photons are diffracted at 45$^\circ$, $k=0$ and the radiation is completely polarized in the direction orthogonal to incidence plane, independently to the polarization of the incident radiation. Since the value of $k$ changes quickly with the diffraction angle, it must be very tight limited around 45$^\circ$ to avoid the reduction of the mean polarization of output radiation. This reduces the diffracted intensity and hence changing the degree of collimation allows to reach a trade-off between the degree of polarization and reasonable fluxes.

The integrated reflectivity, i.e. the efficiency of diffraction, is very low when continuum radiation is used. Since the Bragg diffraction must be satisfied within some eV, a tight constrain on the diffraction angle implies a tight limit in the energy range of photons that can be diffracted. This issue can be overcome with the employment of line emission at energy close to the Bragg energy for 45 degrees diffraction. In this case, incoming monochromatic photons are diffracted with about 50\% efficiency and, even more important, the precise diffraction angle, and hence their degree of polarization, can be easily derived from eq.~\ref{eq:BraggLaw} and eq.~\ref{eq:PolarizationDegree} respectively, thanks to the knowledge of the energy of incoming radiation. In tab.~\ref{tab:Crystals} we report some fluorescence lines which are well tuned with 45$^\circ$ diffraction on several crystals.

\begin{table}[htbp]
\begin{center}
\caption{Tuning between fluorescence lines and diffracting crystals. $\theta$ is the angle of diffraction and ${\mathcal P}$ the polarization of diffracted photons. Data from calculation performed by Henke et al.\cite{Henke1993}.} \label{tab:Crystals}
\begin{tabular}{c|c|c|c|c} 
Line                                           & Energy (keV)  & Crystal                              & $\theta$          & ${\mathcal P}$    \\
\hline
\hline
L$\alpha$ Molybdenum            & 2.293               & Rhodium (001)                 & 45$^\circ$.36  & 0.9994                 \\
K$\alpha$ Chlorine                   & 2.622               & Graphite (002)                  & 44$^\circ$.82   & 0.9986                 \\
L$\alpha$ Rhodium                  & 2.691               & Germanium (111)             & 44$^\circ$.86  & 0.9926                 \\
K$\alpha$ Calcium                    & 3.692               & Aluminum (111)              & 45$^\circ$.88   & 0.9938                 \\
K$\alpha$ Titanium                  & 4.511               & Fluorite CaF$_2$ (220)    & 45$^\circ$.37  & 0.9994                 \\
K$\alpha$ Manganese             & 5.899               & Lithium Floride (220)       & 47$^\circ$.56  & 0.8822                 \\
\end{tabular}
\end{center}
\end{table}

Integrated reflectivity can also be increased with the employment of mosaic crystals, i.e. composed by small domains slightly and regularly misaligned. Each domain acts as an independent flat crystal, but the misalignment allows the diffraction of photons at an angle slightly different from 45$^\circ$ and hence with energy not strictly equal to the Bragg energy at 45$^\circ$.  The employment of mosaic crystals increases the intensity of diffracted radiation, but, when continuum photons are incident, it dilutes the degree of output radiation that can be estimated from its energy width.

\section{The source of polarized photons} \label{sec:Source}

We built a modular source, such that each component is mounted in an aluminum holder, to allow a secure handling and a comfortable change. Indeed the choice of the diffracting crystal allows for the production of polarized photons at different energies, while the selection of collimators allows to reach a reasonable trade-off between diffracted flux and degree of polarization.

\begin{figure}[htbp]
\begin{center}
\subfigure[\label{fig:Capillary}]{\includegraphics[angle=0,totalheight=4cm]{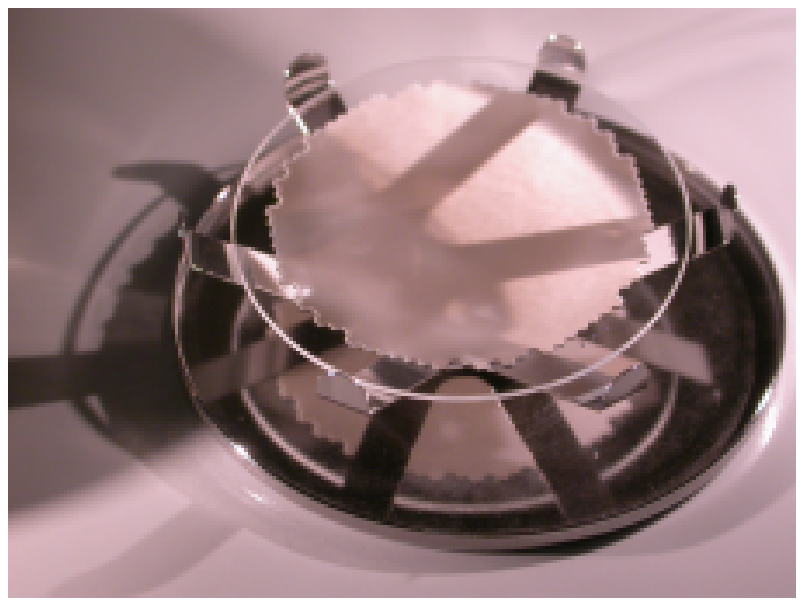}}
\subfigure[\label{fig:CP_holder}]{\includegraphics[angle=0,totalheight=4cm]{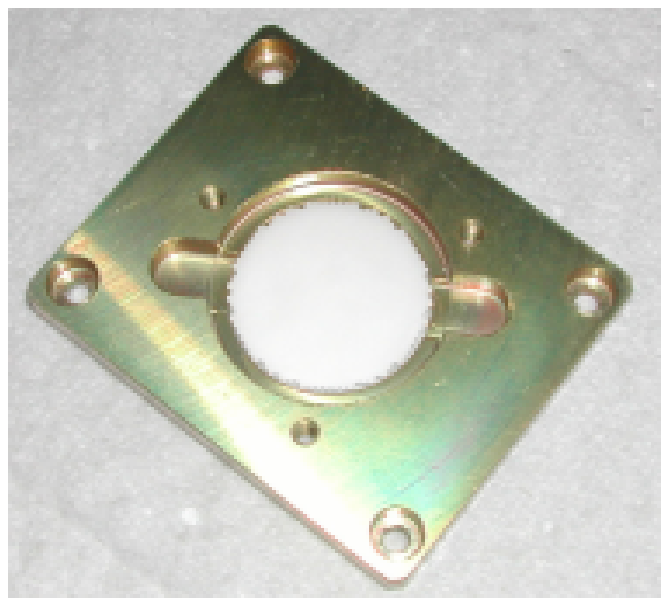}}
\end{center}
\caption{\small ({\bf a}) The lead-glass capillary plate, built by Hamamatsu Photonics, employed as collimator. ({\bf b}) The capillary plate mounted on its aluminum holder. The grooves on the side allow the uplift of the capillary plate with a proper pincer without touching its surface. The ring-shaped cover, fixed with three screws, is not shown.}
\end{figure}

An high degree of collimation, whose importance was outlined in sec.~\ref{sec:Bragg}, and the requirement of a very compact source were achieved with the employment of lead-glass capillary plates (see fig.~\ref{fig:Capillary}). We employed two different types of capillary plates 0.4 and 1.0~mm thick which, thanks to the 10$\mu$m diameter holes, allow good collimation (semi-aperture of $\frac{1}{40}=1^\circ.4$ and $\frac{1}{100}=0^\circ.57$ respectively) with a very limited size. On axis transparency is 57\%, while the effective diameter is 20 and 27~mm for the broad and narrow collimators respectively. Each capillary plate was placed in an aluminum holder (fig.~\ref{fig:CP_holder}) and the external surface was covered with a thin 4~$\mu$m polypropylene film (96\% transparency at 2.6~keV) to protect the small holes by dust.

\begin{figure}[htbp]
\begin{center}
\subfigure[\label{fig:Crystals}]{\includegraphics[angle=0,totalheight=4cm]{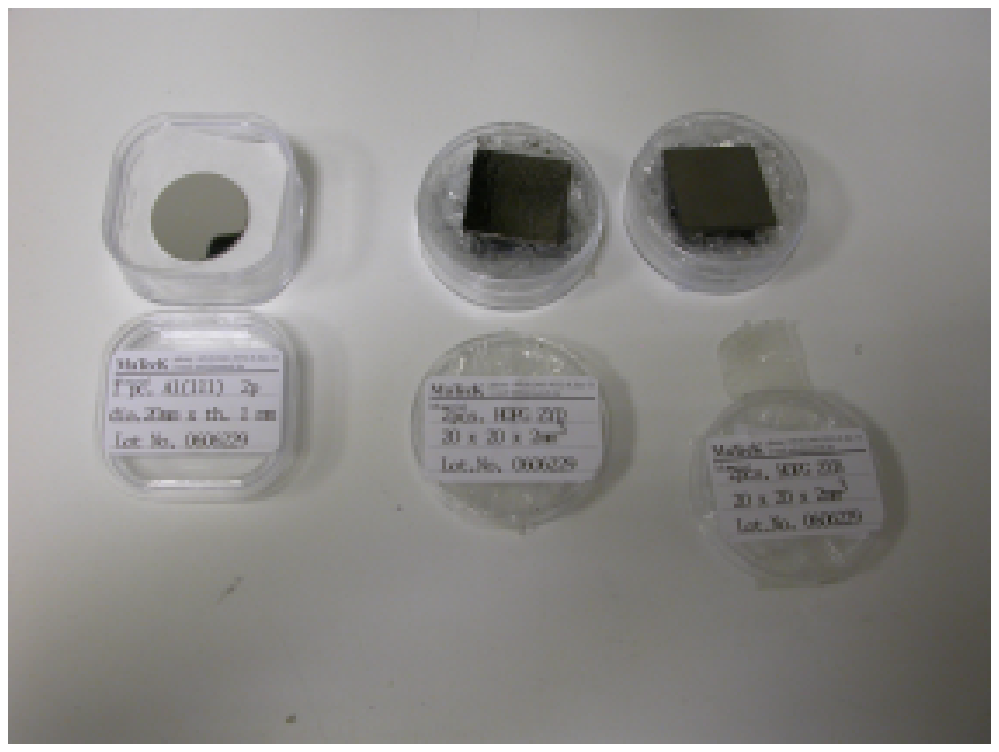}}
\subfigure[\label{fig:Graphite}]{\includegraphics[angle=0,totalheight=4cm]{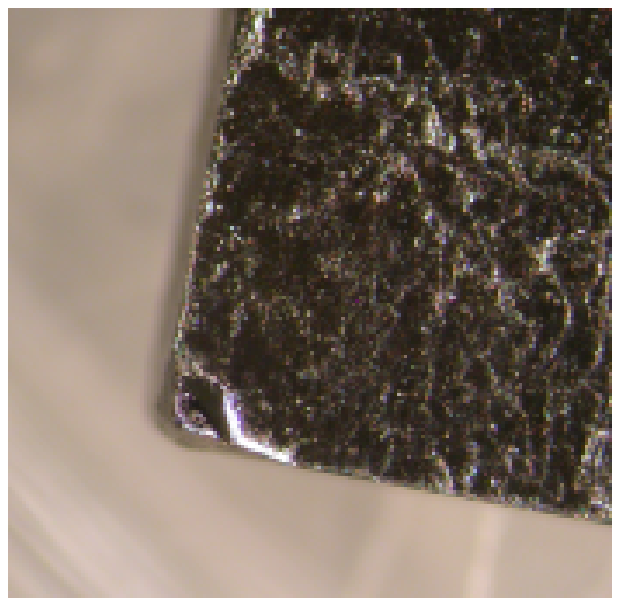}}
\end{center}
\caption{\small ({\bf a}) From left to right: aluminum and graphite crystals built by Matex. Graphite crystals differ from the amplitude of domains misalignment. ({\bf b}) Enlargement view of the grade B graphite crystal, illuminated with grazing incidence light to highlight the roughness of the surface. The edge imperfectness were taken into account in designing the holder.}
\end{figure}

The choice of the crystal allows the selection of the energy of diffracted photons, according to Bragg's Law. We employed graphite mosaic and aluminum flat crystals to produce first (2.6~keV and 3.7~keV) and second (5.2 and 7.4~keV) order polarized photons. Grade B and D $20\times20~\mbox{mm}^2$ graphite crystals were employed, with FWHM gaussian misalignment of 0.8$^\circ$ and 1.2$^\circ$ respectively (fig.~\ref{fig:Graphite}), while aluminum crystal is 20~mm in diameter (fig.~\ref{fig:Crystals}). Crystals also were placed in aluminum holders.

The production of 3.7 keV radiation is especially effective since we employed an X-ray tube with calcium anode as source of unpolarized incident radiation. As reported in table~\ref{tab:Crystals}, the K$\alpha$ line of calcium is very close to the Bragg energy at 45$^\circ$ for flat aluminum crystals and hence it is efficiently diffracted. Polarized photons at 2.6, 5.2 and 7.4~keV are produced by diffraction of continuum radiation, generated with the calcium X-ray tube or with a copper one. Both tubes have very small size but also limited power (200mW).

\begin{figure}[htbp]
\begin{center}
\subfigure[\label{fig:Polarizer_top}]{\includegraphics[angle=0,totalheight=5.5cm]{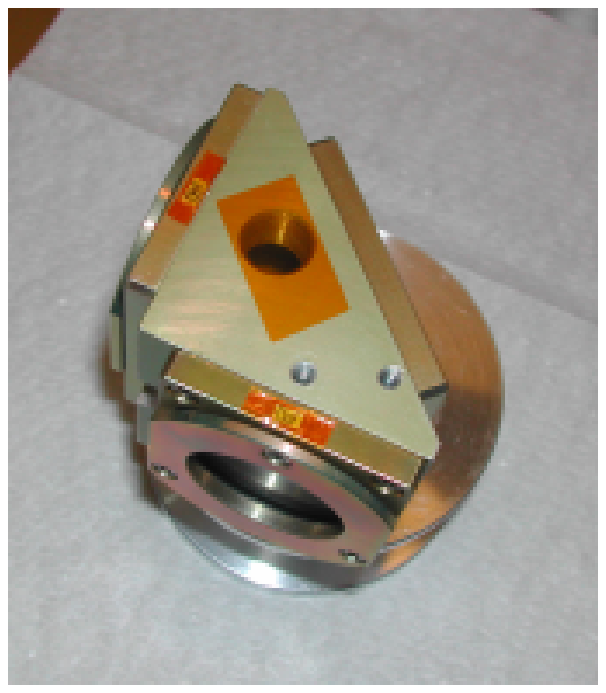}}
\subfigure[\label{fig:Polarizer_side}]{\includegraphics[angle=90,totalheight=5.5cm]{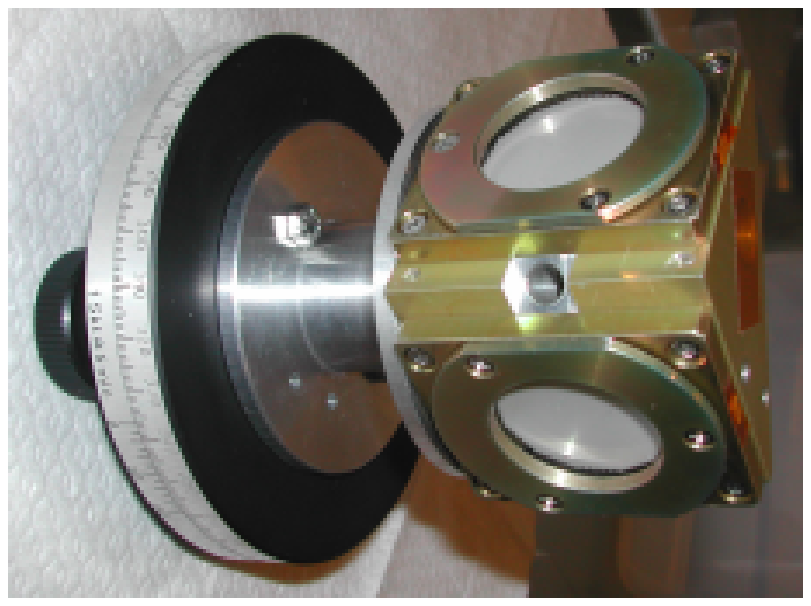}}
\end{center}
\caption{\small ({\bf a}) Top view of the polarizer, showing the capillary plates and the crystal holders mounted on the side of the central half-cube shaped central body. ({\bf b}) The polarizer mounted on a manual rotational stage employed to measure the angle among its components. In the fore the capillary plates are shown.} \label{fig:Polarizer}
\end{figure}

The capillary plates and the crystal holders are mounted on a central half-cube shaped body which constrains the diffraction angle at 45$^\circ$. The so-called polarizer is shown in fig.~\ref{fig:Polarizer}. To reduced air absorption, which heavily affects low energy photons, we limited the distance between capillary plates to about 5~cm. Moreover we add two standard gas connectors, shown in the fig.~\ref{fig:Polarizer_top}, for Helium flowing.

Numerical controlled instruments were employed to built every part of the source to assure the best control on the diffraction angle. Nevertheless we measured the angle between capillary plates and between each capillary and the crystal. The polarizer was mounted on a manual rotational stage and then rotated at 45$^\circ$ steps, measuring the angle  by means of optical laser reflection on capillary plates and aluminum crystal. The crystal holder was aligned to its nominal position within about 0.2$^\circ$ by means of a thin wedge, while the capillary plates resulted aligned.

\begin{figure}[htbp]
\begin{center}
\includegraphics[angle=0,totalheight=5cm]{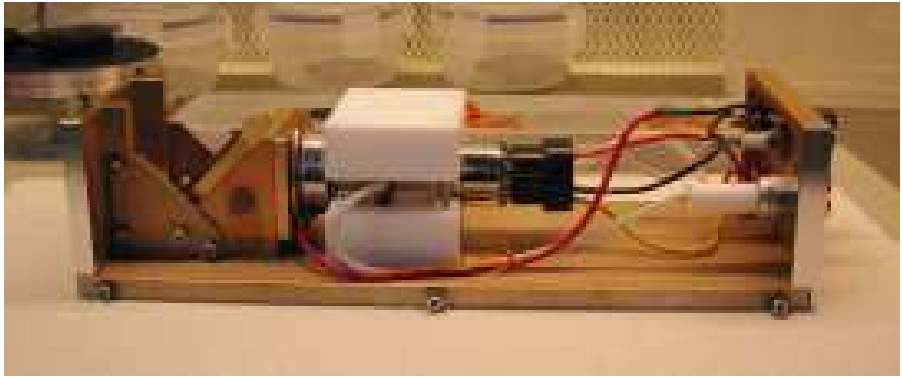}
\end{center}
\caption{\small The complete polarized source. The polarizer is shown on the left, while the X-ray tube is on the right.} \label{fig:Source}
\end{figure}

In fig.~\ref{fig:Source} the complete source is shown. On the right is mounted the polarizer, while the X-ray tube, mounted on an aluminum flange with a teflon support to assure electrical isolation, is on the right. Radiation is downward diffracted. An aluminum case completely cover the source which is $262\times98\times69~\mbox{mm}^3$.

\section{Spectrum and estimated polarization} \label{sec:Performances}

\begin{figure}[htbp]
\begin{center}
\subfigure[\label{fig:FeLine}]{\includegraphics[angle=90,totalheight=4.5cm]{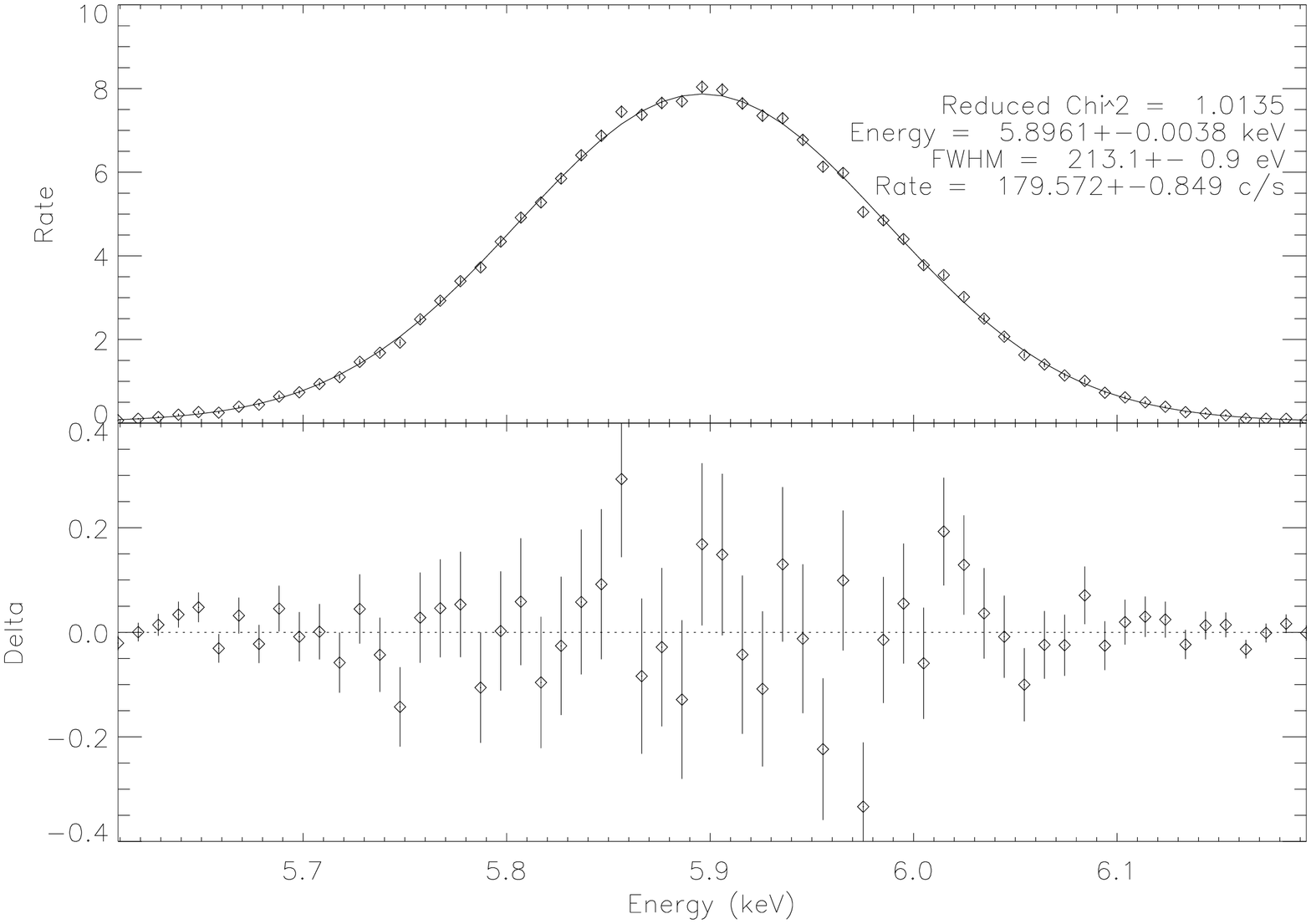}}
\subfigure[\label{fig:Setup}]{\includegraphics[angle=0,totalheight=4.5cm]{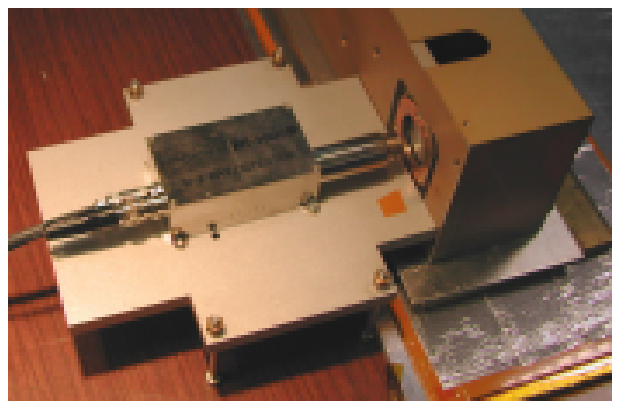}}
\end{center}
\caption{\small ({\bf a}) The spectrum of 5.9~keV line from Fe$^{55}$ radioactive source measured with the Amptek detector. A gaussian fit provides 213~eV FWHM. ({\bf b}) Measurement set-up for the the spectrum of diffracted radiation.}
\end{figure}

We employed an Amptek XR100CR Si-PIN detector, with an energy resolution of 213~eV at 5.9~keV (see fig.~\ref{fig:FeLine}), to measure the diffracted spectrum, since the energy and the width of the diffracted line can be exploited to estimate its degree of polarization. The Amptek detector was placed as close as possible to the output capillary plate to reduced air absorption (see fig.\ref{fig:Setup}). 

\begin{figure}[htbp]
\begin{center}
\subfigure[\label{fig:Gr_I_II}]{\includegraphics[angle=90,totalheight=5cm]{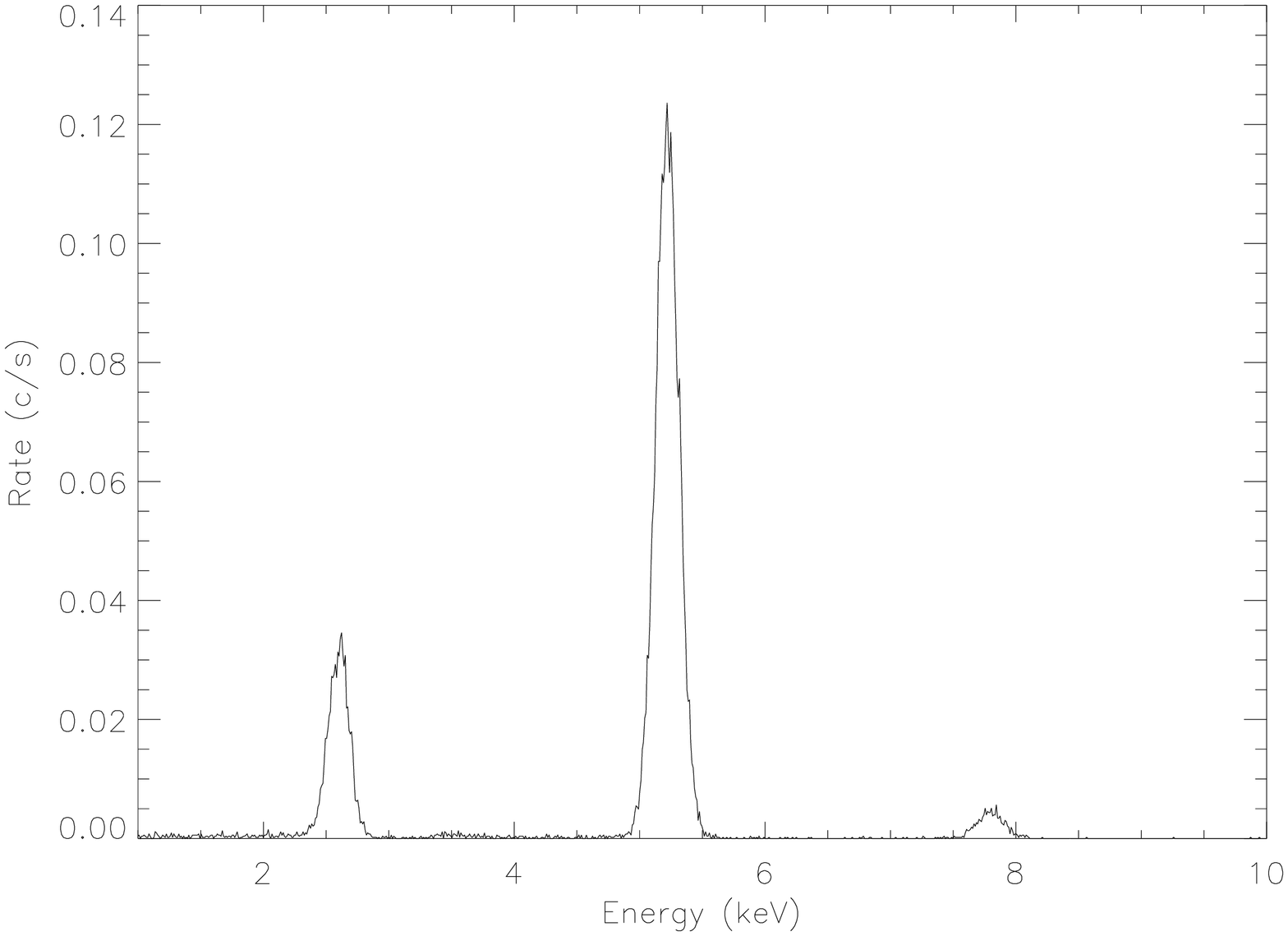}}
\subfigure[\label{fig:Al_I_II}]{\includegraphics[angle=90,totalheight=5cm]{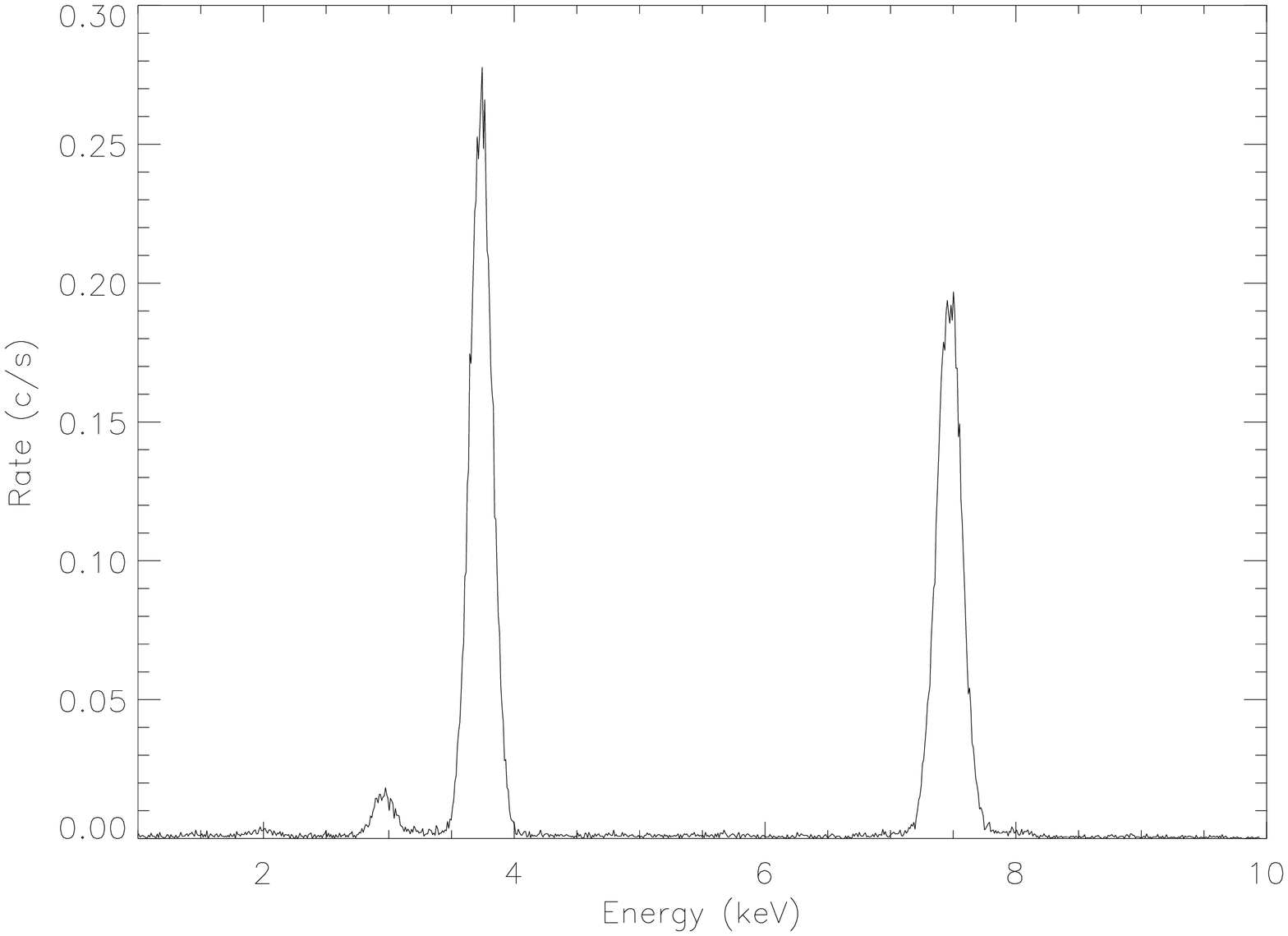}}
\end{center}
\caption{\small The spectrum of radiation diffracted on graphite ({\bf a}) and aluminum ({\bf b}) crystals. In both measurements continuum radiation was employed, hence the relative height depends manly on the incident spectrum and air absorption. Three order of diffraction (2.6, 5.2 and 7.8~keV) are visible in ({\bf a}), while the line at about 3 keV in ({\bf b}) is due to the escape peak in silicon detector and hence is an instrumental artefact.}
\end{figure}

Many configurations, with different collimators, were tested. In fig.~\ref{fig:Gr_I_II} and \ref{fig:Al_I_II} we report the spectrum between 1 and 10~keV for grade~B graphite and aluminum crystal respectively, obtained by means of diffraction of continuum radiation. Two broad capillary plates were employed with the graphite mosaic crystal, while just a single narrow collimator was used for aluminum diffraction. In both spectra very narrow line and low background, manly due to the incomplete charge collection in the Amptek detector, were measured. The relative peaks height depends mainly on the spectrum of the unpolarized incident radiation and on the air absorption. In the spectrum of graphite, the third diffraction order at 7.8~keV is also visible, while the line at about 3~keV in the spectrum of aluminum is due to the escape peak in the silicon detector.


\subsection{Diffraction on graphite}

\begin{figure}[htbp]
\begin{center}
\subfigure[\label{fig:Gr_I}]{\includegraphics[angle=90,totalheight=5cm]{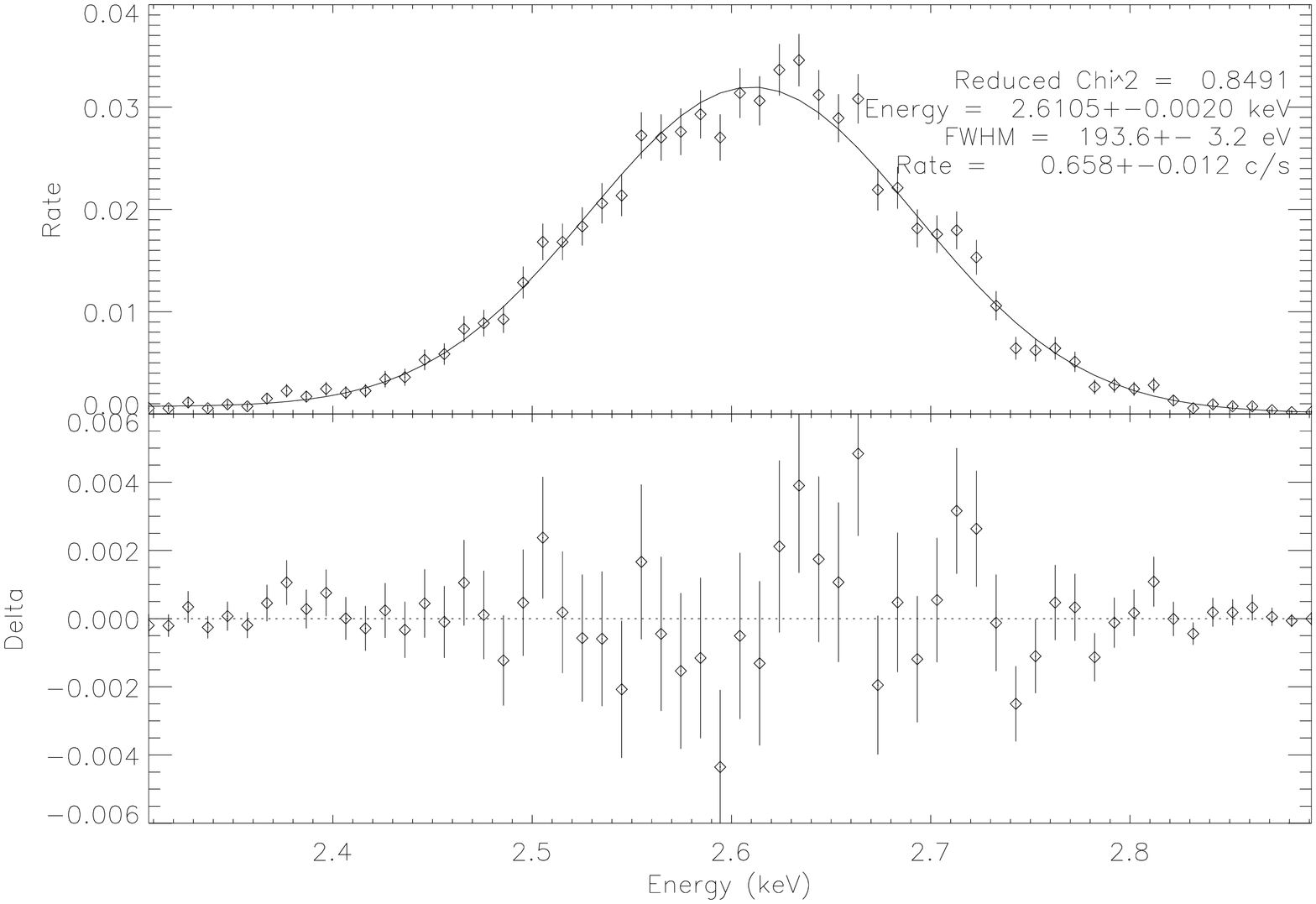}}
\subfigure[\label{fig:Gr_II}]{\includegraphics[angle=90,totalheight=5cm]{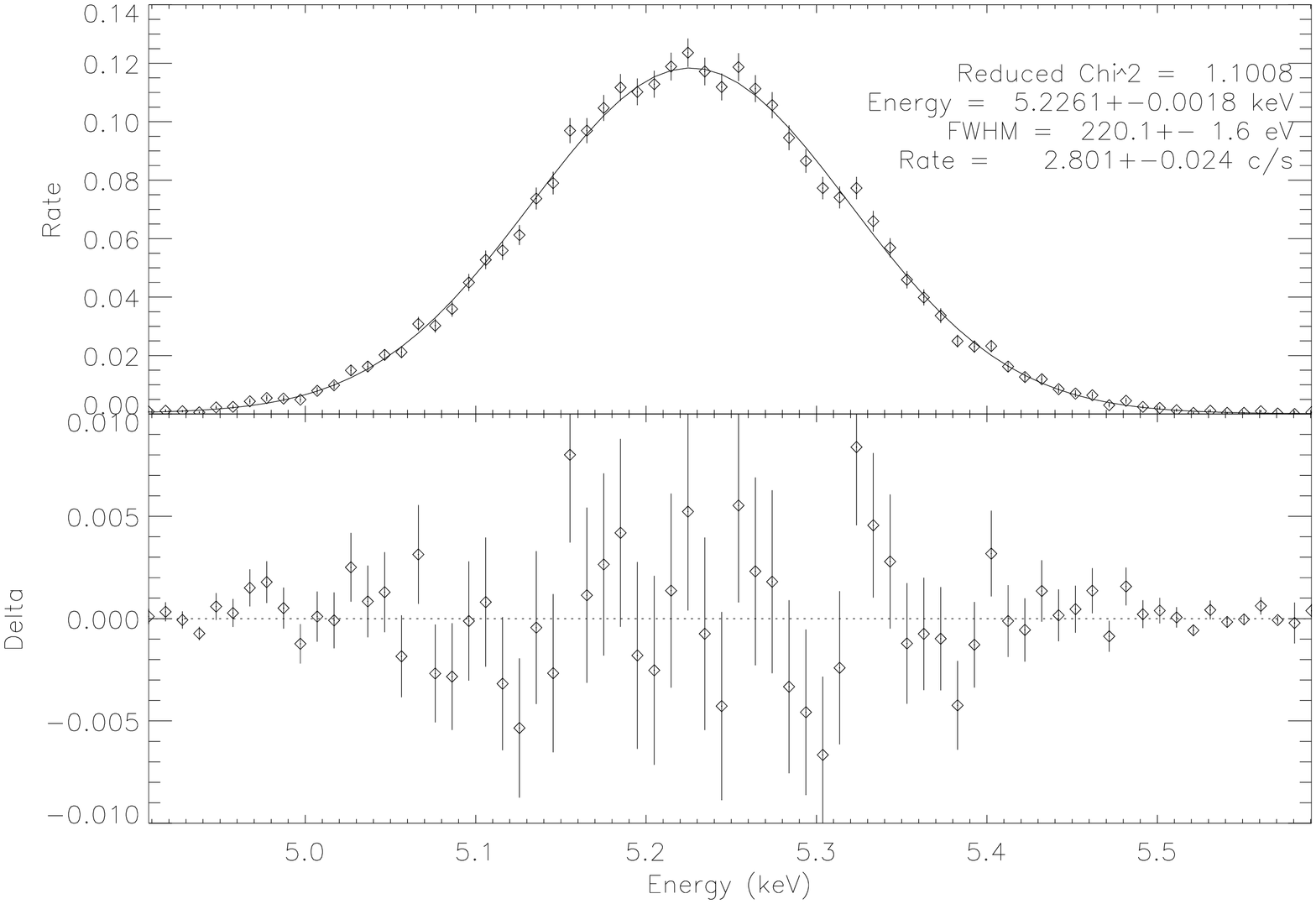}}
\end{center}
\caption{\small Gaussian fit to the spectrum of the first ({\bf a}) and second ({\bf b}) order photons diffracted on graphite grade~B crystals. Two broad collimators, which increase the polarization degree at the expense of the diffracted intensity, were employed.} \label{fig:Gr_I_II_analysis}
\end{figure}

The spectrum of the first and second order diffracted photons on the graphite crystal were fitted with a gaussian profile (see fig.~\ref{fig:Gr_I_II_analysis}). The angle of diffraction, reported in the top part of the table~\ref{tab:Gr_parameters}, is calculated from the measured energy and from Bragg's Law and it is consistent with 45 degrees diffraction within mechanical tolerance. Comparing the line width with the detector resolution, measured by means of fluorescence lines, it results that the first order line is unresolved by Amptek detector and hence an approximate degree of polarization, equal to about 98\%, can be derived from eq.~\ref{eq:PolarizationDegree}. The second order is instead resolved and the 10~eV line width set a 96\% lower limit to the degree of polarization.

A similar analysis was repeated when a single broad collimator was employed. The results, reported in the lower part of the table~\ref{tab:Gr_parameters}, show that a flux increase by a factor 5 can be achieved with a modest decrease of the polarization degree lower limit.

\begin{table}[htbp]
\begin{center}
\caption{Gaussian fit parameters, diffraction angle and estimated polarization for diffracted polarized lines on graphite crystal and different collimation configurations. The increase of the FWHM with energy is partially due to the decreasing resolution of the detector. Air absorption is not taken into account.} \label{tab:Gr_parameters}
\begin{tabular}{|l|c|c|c|c|c|c|c|c} 
\hline
\hline
\multicolumn{6}{l}{Incoming and output $\frac{1}{40}$ collimators} \\
\hline
Diffraction             & Incident radiation & $E$ (keV)                      & FWHM (eV)       & $\chi^2$  & $\theta_{Bragg}$       & $ \mathcal{P}$     & Rate (c/s) \\
\hline
Graphite, I or.  & Continuum           & 2.6105$\pm$0.0020       & 193.6$\pm$3.2  & 0.849         & 45$^\circ$.07            & $\sim0.98$           & 0.66\\
Graphite, II or. & Continuum           & 5.2261$\pm$0.0018       & 220.1$\pm$1.6   & 1.101        & 45$^\circ$.02            & $>0.96$                & 2.8\\
\hline
\hline
\multicolumn{6}{l}{Output $\frac{1}{40}$  collimator} \\
\hline
Graphite, I or. & Continuum            & 2.6109$\pm$0.0036        & 198.6$\pm$4.7   & 1.096       & 45$^\circ$.07           & $>0.95$                & 3.5\\
Graphite, II or. & Continuum           & 5.2269$\pm$0.0037       & 248.2$\pm$2.6    & 0.926       & 45$^\circ$.01            & $>0.94$                & 16.8\\
\hline
\end{tabular}
\end{center}
\end{table}

\subsection{Diffraction on aluminum}

\begin{figure}[htbp]
\begin{center}
\subfigure[\label{fig:Al_I}]{\includegraphics[angle=90,totalheight=5cm]{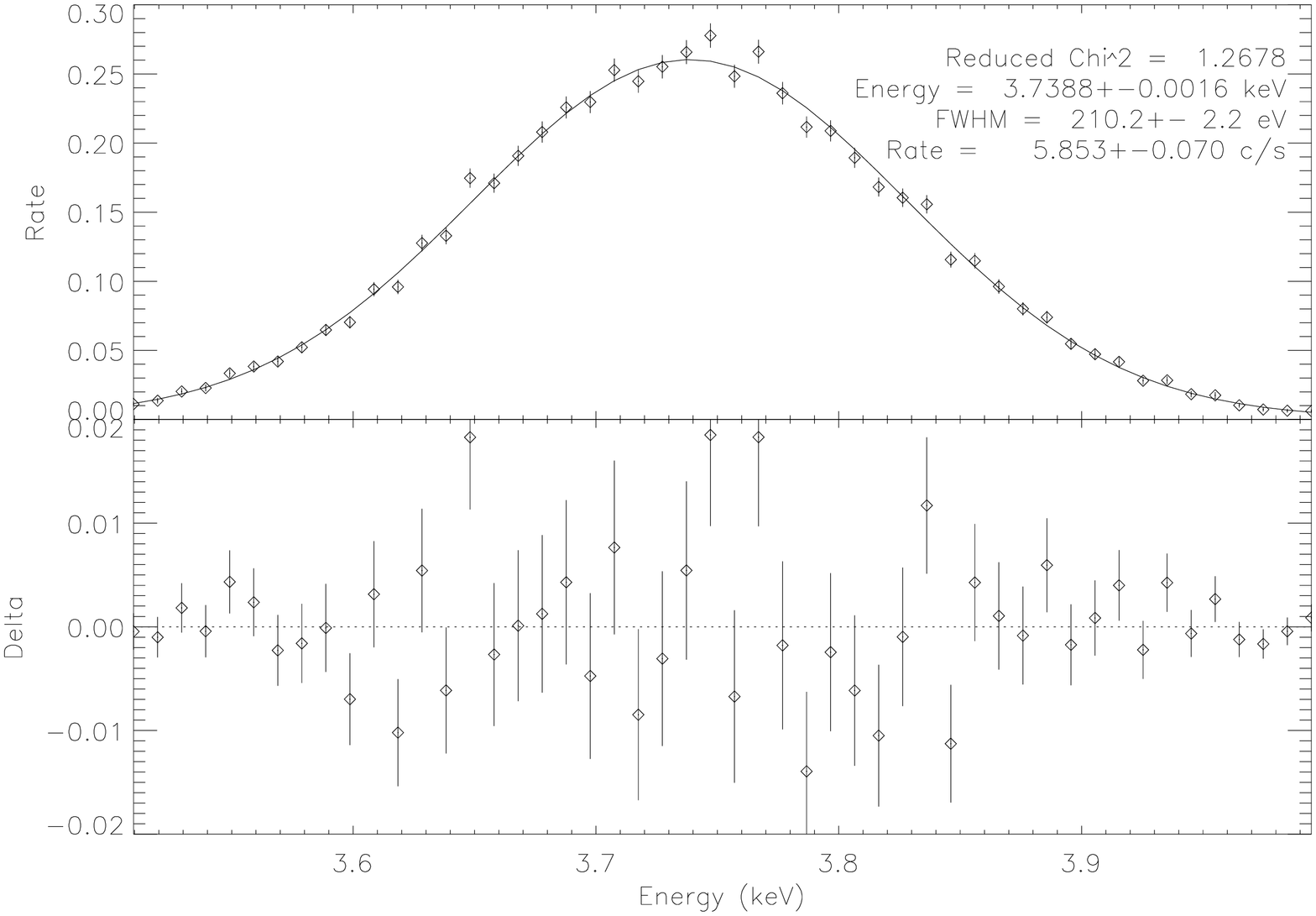}}
\subfigure[\label{fig:Al_II}]{\includegraphics[angle=90,totalheight=5cm]{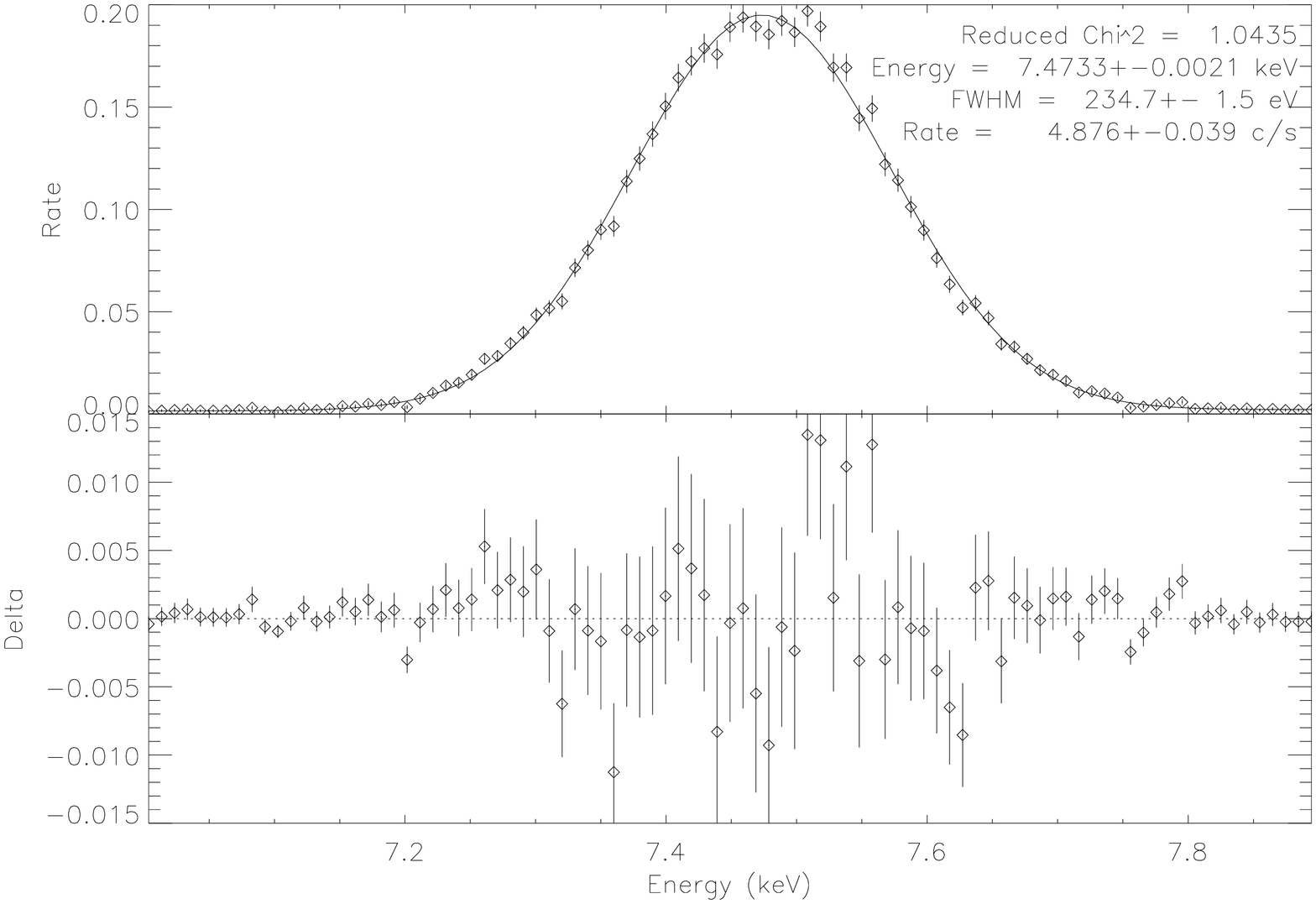}}
\end{center}
\caption{\small Gaussian fit to the spectrum of the first ({\bf a}) and second ({\bf b}) order diffracted radiation on aluminum flat crystals when continuum photons are incident. Only a narrow output collimator is employed.} \label{fig:Al_I_II_analysis}
\end{figure}

The gaussian fit was also applied to the spectrum produced with the aluminum crystal, as shown in fig.~\ref{fig:Al_I_II_analysis}. In the top part of the table~\ref{tab:Al_parameters} we report the fit results when continuum photons, produced by the copper X-ray tube, are diffracted. Since the aluminum crystal is flat, the spectrum is composed by lines which are very narrow and the second order is even unresolved. Hence, very high degree of polarization are inferred.

\begin{figure}[htbp]
\begin{center}
\includegraphics[angle=90,totalheight=5cm]{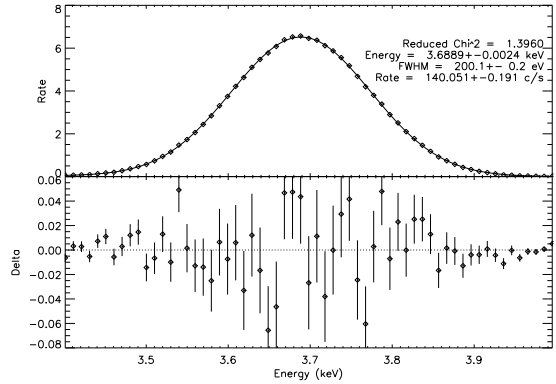}
\end{center}
\caption{\small Gaussian fit to the calcium K$\alpha$ line diffracted on the aluminum flat crystal.} \label{fig:Al_I_line}
\end{figure}

In the lower part of the table~\ref{tab:Al_parameters} we report the analysis performed for the diffraction of radiation produced with calcium X-ray tube. The energy of the line, consistent with the weighted average of the K$\alpha1$ and K$\alpha2$ energy, and the very narrow line width support the lack of any contamination from continuum photons (see fig.~\ref{fig:Al_I_line}). Hence we can assumed that radiation is diffracted at 45$^\circ$.88 (see table~\ref{tab:Crystals}) and is 99.38\% polarized. We stress that the employment of calcium X-ray tube results in a counting rate increase by over a factor 20 respect to the intensity obtained with the diffraction of continuum photons.

\begin{table}[htbp]
\begin{center}
\caption{Gaussian fit parameters, diffraction angle and estimated polarization for diffracted polarized lines on the aluminum crystal. In the top part, continuum emission and a single narrow output collimator is employed. The second order is unresolved by the detector, i.e. $\sim$235~eV FWHM is the resolution of the Amptek detector at that energy. The parameters for diffraction of K$\alpha$ calcium line with a single output broad collimator is reported in the lower part of the table.} \label{tab:Al_parameters}
\begin{tabular}{|l|c|c|c|c|c|c|c|c} 
\hline
\hline
\multicolumn{6}{l}{Output $\frac{1}{100}$ collimators} \\
\hline
Diffraction                   & Incident radiation & $E$ (keV)                      & FWHM (eV)      & $\chi^2$  & $\theta_{Bragg}$       & $ \mathcal{P}$     & Rate (c/s) \\
\hline
Aluminum, I or.   & Continuum           & 3.7388$\pm$0.0016       & 210.2$\pm$2.2   & 1.268         & 45$^\circ$.14            & $>0.97$                & 5.9 \\
Aluminum, II or.   & Continuum          & 7.4733$\pm$0.0021       & 234.7$\pm$1.5   & 1.044        & 45$^\circ$.18             & $\sim0.98$                & 4.9\\
\hline
\hline
\multicolumn{6}{l}{Output $\frac{1}{40}$  collimator} \\
\hline
Aluminum, I or.   & Calcium K$\alpha$ line & 3.6889$\pm$0.0024 & 200.1$\pm$0.2  & 1.396      & 45$^\circ$.93           & $0.9938$              & 140.1\\
\hline
\end{tabular}
\end{center}
\end{table}

\section{The feasibility of on-board calibration source} \label{sec:OnboardSource}

\subsection{The design}

The very compact design of the Bragg polarized source suggests the feasibility of a calibration device for on-board X-ray polarimeters based on a radioactive source. It results from table~\ref{tab:Crystals} that 5.9~keV K$\alpha$ manganese emission, readily produced from Fe$^{55}$ decay, is well in agreement with the 45$^\circ$ Bragg diffraction on lithium floride crystal. However the sensitivity of photoelectric polarimeters such as the Gas Pixel Detector peaks at about 3~keV and hence manganese emission doesn't cover the energy range of major interest. This issue can be overcome by means of 2.6~keV K$\alpha$ chlorine emission, which could be extracted from a thin PVC film irradiated by Fe$^{55}$ source and then polarized with graphite diffraction (see table~\ref{tab:Crystals}). Since the 2.6 and 5.9~keV lines are well resolved in pulse hight by the Gas Pixel Detector, they can provide an effective instrumental calibration in the energy of maximum sensitivity.

\begin{figure}[htbp]
\begin{center}
\includegraphics[angle=90,totalheight=5.cm]{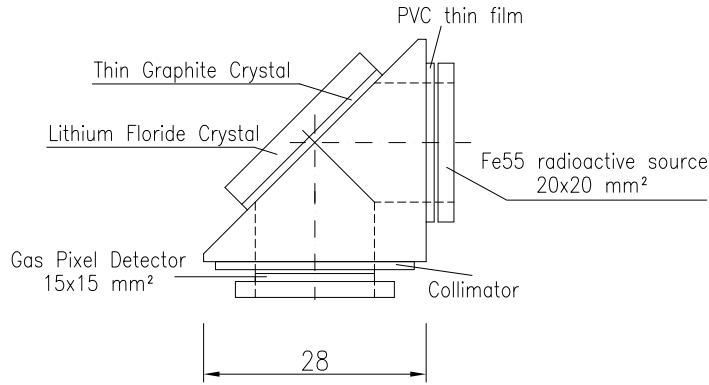}
\end{center}
\caption{\small Sketch of the on-board polarized source.} \label{fig:Sketch}
\end{figure}

In fig.~\ref{fig:Sketch} we show a sketch of the calibration source. A single radioactive Fe$^{55}$ nuclide produces polarized photons at 5.9~keV which are diffracted on a lithium floride crystal and partially absorbed in a thin PVC film, manly from chlorine atoms, and isotropically re-emitted as 2.6~keV fluorescence emission. The latter is diffracted on a thin graphite crystal, evaporated on the lithium floride. If the graphite crystal thickness is 22~$\mu$m, i.e. such that the photon path is equal to the absorption length of 2.6~keV radiation, it results 85\% transparent to 5.9~keV radiation.

Since the 2.6 and 5.9~keV photons are monochromatic, no collimators are required to constrain the diffraction angle at nearly 45$^\circ$. Indeed the Bragg's Law relates the photon energy to the diffraction angle and hence 2.6 and 5.9~keV photons can only be diffracted if they are incident with a grazing angle equal to 44$^\circ$.82 and 47$^\circ$.56  on graphite and lithium fluorite crystal respectively. A broad collimator could be anyhow needed to exclude stray radiation. Neglecting the scattered radiation, monochromatic photons at manly 2.6, 5.9 and 6.5~keV, the latter produced by K$\beta$ manganese emission from  the Fe$^{55}$ decay, are isotropically distributed. Since diffraction on graphite or lithium floride crystals can occur only at the grazing angle reported in table~\ref{tab:Angles}, it results that a collimator is required to stop diffraction on lithium floride of  6.5~keV photons, which occurs at an angle close to 45$^\circ$. Hence we are planning the employment of a 4$^\circ$ aperture collimator centered on 46$^\circ$ angle of diffraction, which stops 6.5~keV emission but allows the transmission of 2.6 and 5.9~keV photons diffracted at 44.8$^\circ$ and 47$^\circ$.6 respectively. A custom capillary plates can be used to provide such collimation with very limited size.

\begin{table}[htbp]
\begin{center}
\caption{Diffraction angle for monochromatic photons produced by Fe$^{55}$ decay and chlorine fluorescence on graphite and lithium floride crystals. The diffraction of chlorine fluorescence lines on the lithium floride crystal is not possible. The production of K$\beta$ chlorine emission is strongly suppressed respect to the K$\alpha$ and hence it is reported only for completeness.} \label{tab:Angles}
\begin{tabular}{r|c|c}
                                                             & Graphite crystal    & Lithium Floride crystal     \\
\hline
2.62~keV K$\alpha$ Chlorine         &  44$^\circ$.8           &  ---                                         \\
2.81~keV K$\beta$ Chlorine           &  41$^\circ$.1           &  ---                                         \\
5.90~keV K$\alpha$ Manganese  &  18$^\circ$.3            &  47$^\circ$.6                      \\
6.49~keV K$\beta$ Manganese     &  16$^\circ$.6            &  42$^\circ$.1                     \\
\end{tabular}
\end{center}
\end{table}

This calibration source, thanks to its limited $28\times28\times28~\mbox{mm}^3$ volume, could be easily placed in a filter wheel so that photons are diffracted orthogonally to the detector plane, allowing the calibration with 99.86\% and 88.22\% polarized lines at 2.6 and 5.9~keV respectively, i.e. in the energy range of major interest for the current design of the Gas Pixel Detector. The entire $15\times15~\mbox{mm}^2$ surface of the detector could be calibrated at the same time with a single Fe$^{55}$ radioactive source with the same effective surface.

\subsection{The PVC thickness}

The thickness of the PVC film should allow a good transparency for the 5.9~keV photons, to perform the high energy calibration, but also it should maximize the production of chlorine fluorescence emission. The optimal thickness can be derived by balancing the absorption of 5.9 and 6.5~keV photons in the PVC and the self-transparency to 2.6~keV radiation.

For a monochromatic parallel beam of radiation orthogonally incident to the PVC film, the transfer equation at energy $E$ can be written\cite{Rybicki1979}:
\begin{equation}
I_E (L) = I_E (0) e^{-\rho \mu_E L} + \int_0^L e^{-\rho \mu_E \left( L-x\right)} j_E(x) dx,
\end{equation}
where $I(L)$ is the beam intensity at a depth $L$, $\rho=1.5 \frac{g}{cm^3}$ is the PVC density, $\mu_E$ is the absorption coefficient at the energy $E$ and $j_E(x)$ is the emission coefficient. Since we are interested in the chlorine 2.6~keV fluorescence emission, we have:
\begin{equation}
I_{2.6} (L) =  \int_0^L e^{-\rho \mu_{2.6} \left( L-x\right)} j_{2.6}(x) dx.\label{eq:TE}
\end{equation}
The emission coefficient is the fluorescence production at a depth $x$ and it is directly proportional to the absorbed photons. Assuming that incident photons have energy equal to  5.9~keV\footnote{The contribution of 6.5~keV photons to the fluorescence emission is about a factor 10 lower.}, the dependence of the emissivity on $x$ can be expressed with:
\begin{equation}
j_{2.6}(x) \propto e^{-\rho \mu_{5.9} x}. \label{eq:Emissivity}
\end{equation}
Substituting the eq.~\ref{eq:Emissivity} in the eq.~\ref{eq:TE} we eventually obtain:
\begin{equation}
I_{2.6} (L) \propto \frac{e^{-\rho\mu_{5.9} L} - e^{-\rho\mu_{2.6} L}} {\rho \left( \mu_{2.6} - \mu_{5.9} \right)}.
\end{equation}
The function $I_{2.6} (L)$ peaks at:
\begin{equation}
L_w = \frac{1}{\rho \left( \mu_{2.6} - \mu_{5.9} \right)} \ln \left( \frac{\mu_{2.6}}{\mu_{5.9}} \right) \approx 41 \mu \mbox{m}
\end{equation}

\begin{figure}[htbp]
\begin{center}
\includegraphics[angle=90,totalheight=5cm]{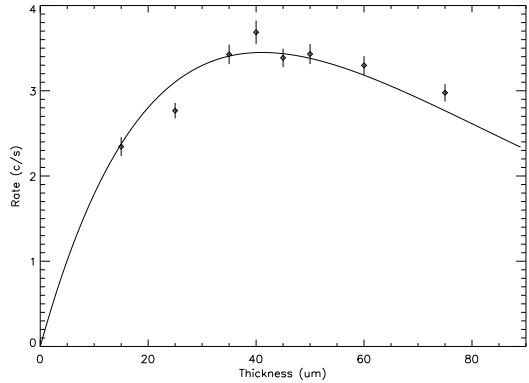}
\end{center}
\caption{\small Measured counting rate from chlorine K$\alpha$ fluorescence emission with respect to the PVC thickness. Only the normalization of the fitting function (eq.~\ref{eq:FittingFunction}) is a free parameter. A 40~$\mu$Ci Fe$^{55}$ source was employed to extract the fluorescence emission.} \label{fig:Fluorescence}
\end{figure}

In fig.~\ref{fig:Fluorescence} we report the measured counting rate of the chlorine 2.6 fluorescence line, extracted with a 40~$\mu$Ci Fe$^{55}$ source, in function of the PVC thickness and fitted with the function:
\begin{equation}
I_{2.6} (x) = K \left[e^{-\rho\mu_{5.9} L} - e^{-\rho\mu_{2.6} L} \right], \label{eq:FittingFunction}
\end{equation}
where only the normalization constant $K$ is a free parameter. The good agreement between expected and measured counting rate confirms that the fluorescence from chlorine is maximized with a $\sim40\mu$m thick PVC film. Such a thin film is 93\% transparent to 5.9~keV radiation, hence allowing also an efficient high energy calibration.

\subsection{The intensity of the Fe$^{55}$ radioactive source}

\begin{figure}[htbp]
\begin{center}
\subfigure[\label{fig:Diffraz_Fe55}]{\includegraphics[angle=0,totalheight=5cm]{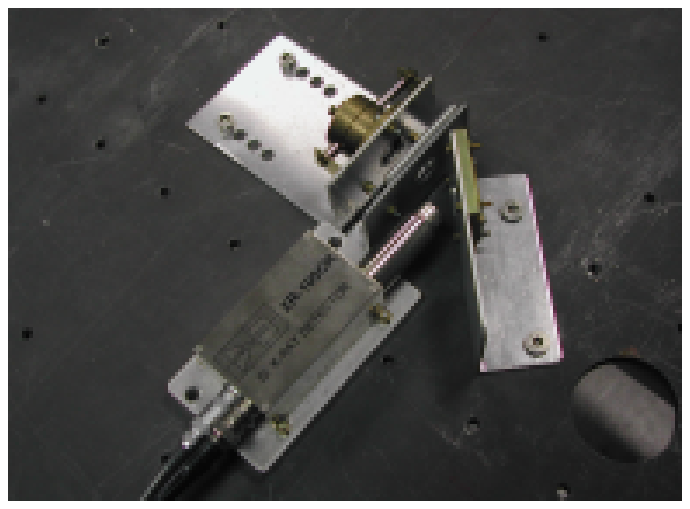}}
\subfigure[\label{fig:SpectrumDiffraz}]{\includegraphics[angle=90,totalheight=5cm]{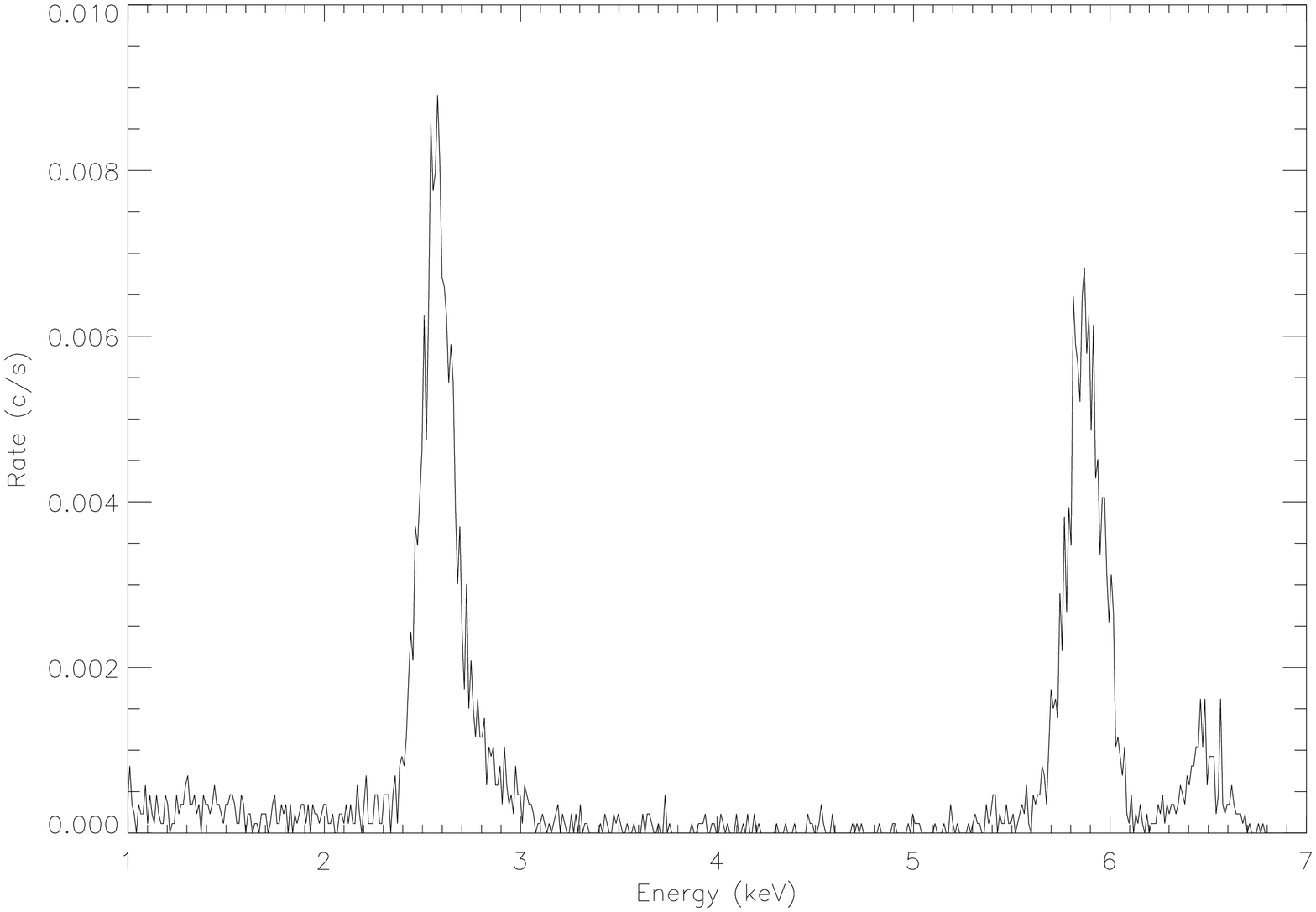}}
\end{center}
\caption{\small ({\bf a}) Set-up for the study of the diffracted K$\alpha$ chlorine line intensity. The Fe$^{55}$ source is on the top left, in the brass holder, while the graphite crystal is on the right. ({\bf b}) Spectrum of the diffracted emission when a  40~$\mu$Ci Fe$^{55}$ is employed. The fluorescence of manganese at 5.9~keV and 6.5~keV is scattered and not diffracted, even if an high polarization is expected in any case.}
\end{figure}

To estimate the intensity of the Fe$^{55}$ source required to achieve a reasonable flux on the Gas Pixel Detector, we measured the diffracted intensity when the 40~$\mu$Ci Fe$^{55}$ source is employed. The source, placed in a brass holder on a optical table, extracts chlorine fluorescence emission from a 45~$\mu$m PVC film that is diffracted on a grade~B graphite crystal (see fig.~\ref{fig:Diffraz_Fe55}). No collimator was employed. 

The spectrum (see fig.~\ref{fig:SpectrumDiffraz}) was measured with the Amptek detector and it is composed by the diffracted radiation at 2.6~keV and from scattered 5.9 and 6.5~keV emission from Fe$^{55}$. We expect that scattered components, that are only slightly less intense respect to the diffracted one, would be negligible in the on board calibration source, since graphite thickness would be reduced to some tens of micron, i.e. it would be almost transparent to the 5.9 and 6.5~keV photons. 

A gaussian fit to the line at 2.6~keV returns a counting rate of about 0.13~counts/s. Accounting for the air absorption, the Amptek detector $5\times5~\mbox{mm}^2$ area and for the efficiency of the Gas Pixel Detector filled with the standard He~20\% DME~80\% mixture, we expect about 110~counts/s for a 5~mCi source. Since the decay time of the Fe$^{55}$ is about 2.7~years, a moderately strong source is sufficient for the calibration of the Gas Pixel Detector for the whole length of a dedicated space borne mission.

\section{Conclusion} \label{sec:Conclusion}

We presented a source, based on the Bragg diffraction, which can provide nearly monochromatic and completely polarized photons. Unpolarized radiation, produced with an X-ray tube, is incident at nearly 45$^\circ$ on a crystal and, since only photons polarized orthogonally to the diffraction plane and with energy equal to the Bragg energy at 45$^\circ$ are efficiently diffracted, the output radiation is highly polarized and nearly monochromatic. The choice of the diffracting crystal allows the production of polarized photons at different energies. Thanks to the employment of capillary plates which can constrain the diffraction angle to nearly 45$^\circ$ with limited size and to small power X-ray tubes, the source is also very compact (it is enclosed in a $262\times98\times69~\mbox{mm}^3$ box) and transportable. 

In particular, we presented the production of 2.6, 5.2 and 3.7, 7.4~keV photons exploiting the first and second order diffraction on graphite and aluminum crystals respectively. Incident photons are generated with copper and calcium X-ray tubes. The latter is particularly effective when employed with the aluminum crystal, since the calcium K$\alpha$ line is diffracted at nearly 45$^\circ$ and hence the output radiation results much more intense than that achieved with diffraction of continuum photons.

We also presented a feasibility study for the construction of a source, based on the Bragg diffraction and on a Fe$^{55}$ radioactive source, which, thanks to its very limited size (about $28\times28\times28~\mbox{mm}^3$), could be employed for the calibration of the next generation X-ray polarimeters, such as the Gas Pixel Detector. A Fe$^{55}$ nuclide could be used to extract 2.6~keV fluorescence emission from chlorine atoms, contained in a PVC thin film, and to produced 5.9~keV emission. Photons at 2.6 and 5.9~keV are then diffracted on a graphite and lithium floride crystals respectively, producing polarized radiation in the energy range of major interest for the employment of the Gas Pixel Detector.

\section*{Acknowledgments}

This research is supported by ASI contract I/074/06/0 and INAF PRIN 2005. We also thank the Venfin (Rho, Italy) to kindly provide us the thin PVC films which was employed for the study presented in sec.~\ref{sec:OnboardSource}.

\bibliography{References}   
\bibliographystyle{spiebib}   

\end{document}